\documentclass{IEEEtran}
\def\BibTeX{{\rm B\kern-.05em{\sc i\kern-.025em b}\kern-.08em
    T\kern-.1667em\lower.7ex\hbox{E}\kern-.125emX}}

\usepackage{graphicx}
\usepackage{xcolor}
\usepackage{comment}
\usepackage{amsmath,amssymb,amsfonts}
\usepackage{algorithmic}
\usepackage{graphicx}
\usepackage{textcomp}
\usepackage{comment}
\usepackage{mathtools, cuted}
\usepackage{relsize}
\usepackage{comment}
\usepackage{cite}

\usepackage{algorithm}
\usepackage{multicol}
\usepackage[normalem]{ulem}

\def\rev#1{{#1}}

\def\CD#1{\textcolor{red}{#1}}

\def\CDnote#1{\textcolor{red}{\textbf{[CD $\bullet$} \textit{#1}\textbf{]}}}

\def\GC#1{\textcolor{blue}{#1}}

\newtheorem{remark}{Remark}
\newtheorem{lemma}{Lemma}
\newtheorem{theorem}{Theorem}
\newtheorem{definition}{Definition}
\newtheorem{proposition}{Proposition}
\newtheorem{assumption}{Assumption}

\begin{document}
\title{A kernel-based approach to physics-informed nonlinear system identification$^\S$}
\author{Cesare Donati$^{1,2,\star}$, \IEEEmembership{Graduate Student Member, IEEE}, Martina Mammarella$^{2}$, \IEEEmembership{Senior Member, IEEE},\\ Giuseppe C. Calafiore$^1$, \IEEEmembership{Fellow, IEEE}, Fabrizio Dabbene$^{2}$, \IEEEmembership{Senior Member, IEEE},\\  Constantino Lagoa$^{3}$, \IEEEmembership{Member, IEEE}, and Carlo Novara$^1$, \IEEEmembership{Senior Member, IEEE}
\thanks{$^\S$ This work has been submitted to the IEEE for possible publication.
Copyright may be transferred without notice, after which this version may
no longer be accessible.}
\thanks{$^{\star}${Corresponding author: { cesare.donati@polito.it}.}}
\thanks{$^{1}$DET, Politecnico di Torino, Corso Duca degli Abruzzi 24, Torino, Italy.} %{cesare.donati@polito.it, carlo.novara@polito.it}.}%
\thanks{$^{2}$CNR-IEIIT, c/o Politecnico di Torino, Corso Duca degli Abruzzi 24, Torino, Italy.}
%, {martina.mammarella@cnr.it, fabrizio.dabbene@cnr.it}.}%
\thanks{$^{3}$EECS, The Pennsylvania State University, University Park, PA, USA.}
%, {cml18@psu.edu}.}%
}

\maketitle
%\thispagestyle{empty}
%\pagestyle{empty}

%%%%%%%%%%%%%%%%%%%%%%%%%%%%%%%%%%%%%%%%%%%%%%%%%%%%%%%%%%%%%%%%%%%%%%%%%%%%%%%%
\begin{abstract}
This paper presents a kernel-based framework for physics-informed nonlinear system identification.
The key contribution is a structured methodology that
extends kernel-based techniques to seamlessly embed partially
known physics-based models, improving parameter estimation
and overall model accuracy.
The proposed method enhances traditional modeling approaches by embedding a parametric model, which provides physical interpretability, with a kernel-based function,
which accounts for unmodeled dynamics. 
The two models' components are identified from the data simultaneously, thereby minimizing a suitable cost that balances the relative importance of the physical and the black-box parts of the model.
Additionally, nonlinear state smoothing is employed to address scenarios involving state-space models with not fully measurable states. Numerical simulations on an experimental benchmark system demonstrate the effectiveness
of the proposed approach, 
\rev{achieving up to $51\%$ reduction in simulation root mean square error compared to physics-only models and $31\%$ performance improvement over state-of-the-art identification techniques.}
%, with performance comparisons validating the superior accuracy of the integrated kernel-physics framework.
%with performance comparisons against state-of-the-art identification techniques.
\end{abstract}
%\begin{keywords}
%Nonlinear systems identification,  Grey-box modeling, Estimation, Kernel methods.
%\end{keywords}

%%%%%%%%%%%%%%%%%%%%%%%%%%%%%%%%%%%%%%%%%%%%%%%%%%%%%%%%%%%%%%%%%%%%%%%%%%%%%%%%
\section{Introduction}
Nonlinear system identification plays a crucial role in engineering, aiming to construct models that accurately represent the complex dynamics of physical systems using measured data~\cite{ljung2019CSM}. Traditional approaches often rely on parametric models derived from physical principles (also referred to as {off-white} models) \cite{ljung2010perspectives} or more flexible representations (black-box models), such as neural networks \cite{mavkov2020integrated} or nonlinear basis function combinations \cite{svensson2017flexible}. Recently, the combination of these two approaches has been proposed to tackle the challenge of compensating for unmodeled dynamics in physical models that arise in real-world applications (see, e.g., \cite{quaghebeur2021incorporating,AutomaticaExtendedVersion}). 

{Within this context, some identification approaches follow a two-step process: first, they estimate the physical parameters while assuming no unmodeled dynamics, and then they introduce corrections to model the resulting \textit{discrepancy} \cite{brynjarsdottir2014learning,kaheman2019learning,forgione2023adaptation}. This strategy inevitably produces biased physical parameter estimates, which need to be handled a posteriori {by compensating them via the black-box component
of the model. Such a term, therefore, must not only account for modeling errors but also compensate for the bias error induced by the parametric model identification phase.}
An alternative perspective is presented in, e.g., \cite{zhu2022robust,AutomaticaExtendedVersion}. By modeling unmodeled dynamics explicitly from the beginning and estimating both the physical parameters and correction terms simultaneously, the interference between the two is minimized, leading to a more accurate and reliable identification process that prevents biased parameter estimates. In this context, sparsification is essential to ensure that corrections remain interpretable and do not overshadow the underlying physical model. However, despite their effectiveness in identifying nonlinear models, a key limitation of these methods is the need for a careful selection of appropriate basis functions that can adequately capture the underlying unmodeled system dynamics \cite{care2023kernel}.
%Kernel methods naturally provide a powerful framework for overcoming these challenges, making them particularly well-suited for the problem at hand.
}

Kernel methods \cite{pillonetto2014kernel,care2023kernel} are a class of nonparametric machine learning techniques, able to provide a powerful framework for overcoming these challenges by enabling the construction of regularized models directly from data, without the need for explicitly defining basis functions. These methods are widely used in the context of input-output system identification (see, e.g., \cite{schoukens2016modeling,dalla2021kernel}), where the relationship between inputs and outputs is learned directly from measured data. However, the conventional kernel approaches do not incorporate physical system knowledge, which on the other hand can be crucial for developing interpretable and reliable models, {featuring also improved generalization capabilities.} %in many engineering applications.

In this work, we aim to fill this gap by proposing a novel identification framework that embeds kernel methods with available physics-based models. The approach leverages kernel-based function approximation to systematically compensate for unmodeled dynamics, while preserving the interpretability of the physical part of the model. 
Unlike traditional methods that rely on predefined, often heuristically chosen, basis function dictionaries, the proposed formulation adapts directly to the data. Thus, by leveraging the representer theorem~\cite{scholkopf2001.reprthm}, the proposed framework provides a regularized, data-driven functional approximation mechanism that eliminates the need for manual selection of the basis functions while preserving both interpretability and accuracy through the embedded physical structure.

%\MM{Moreover, to encompass a broader class of dynamical systems \cite{ribeiro2020smoothness}, we rely on a more general state-space setting, unlike traditional input-output identification models (e.g., NARX).} 
{Moreover, to encompass a broader class of dynamical systems \cite{ribeiro2020smoothness}, we rely on a more general state-space setting, extending the proposed method beyond traditional input-output identification models (e.g., NARX).}
%Moreover, we extend the proposed method beyond traditional input-output identification models, e.g., (nonlinear) autoregressive models with exogenous inputs (NARX), to a more general state-space setting, which encompasses a broader class of dynamical systems \cite{ribeiro2020smoothness}. 
{In fact, many physical systems are better described by state-space models, which explicitly capture system dynamics over time \cite{care2023kernel}. This formulation also unifies various prediction models, including nonlinear output error, ARMAX, and ARX models \cite{ribeiro2020smoothness}, and serves as a foundation for numerous controller and observer design methods, making it particularly valuable for system identification.} %Additionally, many controller and observer design approaches are based on a state-space representation. 

{Unlike static regression models, a state-space formulation accounts for the evolution of hidden states, requiring estimation of both system parameters and unmeasured state trajectories.
One common approach to this challenge is multi-step identification, where unmeasured states are recursively estimated through repeated model evaluation (see, e.g., \cite{farina2011simulation,terzi2018learning,AutomaticaExtendedVersion}). While this strategy naturally estimates latent states by iterating the estimation model, it often makes the optimization challenging due to its strong nonlinear parameter dependencies. Moreover, extending the kernel-based framework to multi-step settings introduces further complexity. To circumvent these issues, we adopt an alternative strategy inspired by \cite{schon2011system,frigola2013bayesian}, in which prior state estimates, derived from available data, are used within prediction-based state-space optimization problems.
To this end, we combine
%a nonlinear filter based on
an unscented Kalman filter (UKF) \cite{wan2000unscented} with an unscented Rauch–Tung–Striebel smoother (URTSS) \cite{sarkka2008unscented,akhtar2023computationally} to reconstruct the hidden state trajectories, enabling kernel-based model embedding in a state-space setting.}
%\CD{Smoothing techniques have been previously explored in system identification to enhance state estimation and model accuracy, as seen in, e.g., \cite{schon2011system,frigola2013bayesian,jonker2020efficient}. In this work, we present a realization of these techniques within our framework to extend the applicability of the kernel-based approach.} 
%{While smoothing techniques have been used in system identification to improve state estimation and model accuracy, see e.g.,  \cite{schon2011system,frigola2013bayesian,jonker2020efficient}, their integration within kernel-based frameworks remains less explored. 
%Here, we leverage them to {reconstruct the unobservable states, reduce measurement noise effects}, and enable kernel-based model augmentation in a state-space setting.}

The effectiveness of the proposed approach is validated through an experimental benchmark using real data, showcasing its advantages over state-of-the-art techniques in both predictive accuracy and simulation performance.

The remainder of the paper is structured as follows. In Section~\ref{sec:kerneltheory}, we provide an introduction to nonlinear function estimation using kernel theory. The main contribution of this paper, i.e., embedding the parameterized physical models with data-driven kernels to account for unmodeled dynamics, is presented in Section~\ref{sec:augkernel}. Then, in Section~\ref{sec:sskernel} we extend the kernel-based framework to state-space systems via nonlinear state smoothing, enabling its application to a broader class of systems. Finally, numerical results on the experimental benchmark are discussed in Section~\ref{sec:results} and main conclusions are drawn in Section~\ref{sec:concl}.

%\CDnote{Kernel method: \cite{pillonetto2014kernel,care2023kernel}  (survey/tutorial), \cite{pillonetto2010new} (methodological, linear), \cite{pillonetto2011new} (methodological, nonlinear). Representer theorem: \cite{scholkopf2001.reprthm}.}

%\CDnote{1. Classic introduction on NSysId and needs for compensations of unmodeled parts in physical models. 2. Problem of selecting the correct basis function. 3. Kernel methods solve this problem. 4. Kernel methods are applied to input/output, single-step system and do not consider physics. 5. Generality of State-Space system (ability to model different class of systems (ARX, ARMAX, OE) \cite{ribeiro2020smoothness}. 6. In this work: generalization of kernel methods to augment physical model in a single-step state space setting, i.e., a very general approach. }

%\noindent\textit{Notation.} $I_n$ represents the $n\times n$ identity matrix.

%\newpage

\section{Kernel-based approximation}\label{sec:kerneltheory}
In this section, we provide a brief introduction to nonlinear function approximation using kernels, which represents the core foundation for the main results presented in this paper.
First, we introduce two definitions related to kernels.

\begin{definition}[positive definite kernel \cite{scholkopf2001.reprthm}]\label{def:pdkernel} Let $\mathcal X$ be a nonempty set. A real-valued, continuous, symmetric function $\kappa: \mathcal X\times\mathcal X \to \mathbb R$ is a positive-definite kernel (on $\mathcal{X}$) if $\sum_{i=1}^n\sum_{j=1}^n c_ic_j\kappa(x_i,x_j)\ge0$ holds for all $n\in\mathbb N$, $x_1,\dots,x_n\in\mathcal X$, $c_1,\dots,c_n\in\mathbb R$. %\MM{Why $n$ and not $m$?}
\end{definition}
\begin{definition}[reproducing kernel Hilbert space \cite{aronszajn1950theory,wahba1990spline}]\label{def:RKHS}
Let $\mathcal{H}$ be a Hilbert space of real-valued functions defined on a nonempty set $\mathcal{X}$, with inner product $\langle\cdot,\cdot\rangle_\mathcal{H}$. A function $\kappa: \mathcal X\times\mathcal X \to \mathbb R$ is a reproducing kernel of $\mathcal{H}$, and $\mathcal{H}$ is a reproducing kernel Hilbert space (RKHS) on $\mathcal{X}$ if the following conditions hold: (i) For any $x\in\mathcal{X}$, $\kappa(\cdot, x)\in\mathcal{H}$, and (ii) the \textit{reproducing property} holds, i.e., for any $x\in\mathcal{X}$, $h\in\mathcal{H}$, $\langle h(\,\cdot\,),\kappa(\cdot,x)\rangle_\mathcal{H}=h(x)$.
\end{definition}

From the reproducing property in Definition~\ref{def:RKHS}.b, we observe that the value of $h$ in $x$ can be represented as an inner product in the feature space. Hence, applying this property to the kernel $\kappa$, for any $x, x^{\prime} \in \mathcal{X}$, we have that  $\kappa(x,x^\prime) = \langle \kappa(\cdot,x),\kappa(\cdot,x^\prime)\rangle_\mathcal{H}$. Moreover, according to the Moore–Aronszajn theorem \cite{aronszajn1950theory}, we know that every positive definite kernel $\kappa$ uniquely defines a RKHS $\mathcal{H}$ in which $\kappa$ serves as the reproducing kernel. Conversely, each RKHS is associated with a unique positive definite kernel. Common choices for the kernel function $\kappa(x, x^{\prime})$ include the Gaussian (radial basis function) kernel $\kappa(x, x^{\prime}) = \exp(-\|x - x^{\prime}\|^2_2 / 2\sigma^2)$, the Laplacian kernel $\kappa(x, x^{\prime}) = \exp(-\|x - x^{\prime}\|_1 / \sigma)$, the polynomial kernel $\kappa(x, x^{\prime}) = (x^\top x^{\prime} + c)^d$, and the linear kernel $\kappa(x, x^{\prime}) = x^\top P x^{\prime}$, where $\sigma$, $c$, $d$, and $P\succeq0$ are hyperparameters.

{Given a kernel $\kappa$, we next consider a nonlinear input-output relation given by an unknown nonlinear function 
 $ g:\mathcal{X}\to\mathbb{R}$
\begin{equation}
    y = g(x) + e,
    \label{eqn:system}
\end{equation}
where 
$x\in \mathcal X$, $y\in \mathbb R$
are the input and output, respectively, 
 $g$ is assumed to belong to the native RKHS $\mathcal{H}$ associated with the given kernel $\kappa$, and $e \in \mathbb R$ is
 an error term which represents  measurement noise, as well as  possible structural errors on $g$.}
Let $\mathcal{D}=\{(x_1,y_1),\dots,(x_T,y_T)\}$ be a sequence of given $T$ input-output data, collected from the system \eqref{eqn:system}.
The goal is to find an estimate $\hat g$ of the function $g$, accurately representing the observed data while ensuring that, for any new pair of data $(x,y)$, the predicted value $\hat g(x)$ remains close to $y$. 

A standard approach to estimate $g$ using the dataset $\mathcal{D}$ involves minimizing a loss function that combines a quadratic data-fit term (i.e., the prediction error) with a regularization term. Hence, the unknown function $g$ can be estimated by solving the following \rev{well-known kernel ridge regression (KRR) \cite{saunders1998KRR}} problem
\begin{equation}
    \hat g = \arg \min_{g \in \mathcal{H}} \sum_{t=1}^{T} (y_t - g(x_t))^2 + \gamma\|g\|^2_{\mathcal{H}}, 
    \label{eqn:optprob.init}
\end{equation}
where $\gamma\in\mathbb{R}$ is a trade-off weight that balances data-fit and regularization, {and $\|g\|_{\mathcal{H}}=\sqrt{\langle g,g\rangle}_\mathcal{H}$ is the norm in $\mathcal{H}$, introduced by the inner product $\langle \cdot,\cdot\rangle_\mathcal{H}$.} %\textcolor{green}{(It seems that the norm $\| \cdot \|_{\mathcal{H}}$ is not defined. I guess it is the one arising from the inner product...)}
Then, the \emph{representer theorem}~\cite{scholkopf2001.reprthm} guarantees the uniqueness of the solution to~\eqref{eqn:optprob.init}, expressed as a sum of $T$ basis functions determined by the kernel, with their contributions weighted by coefficients obtained through the solution of a system of linear equations (see also, e.g., \cite{wahba1990spline}).
In particular, given a positive-definite, real-valued kernel $\kappa: \mathcal{X}\times\mathcal{X}\to \mathbb R$, and defining the kernel matrix $\mathbf{K} \in \mathbb{R}^{T,T}$ with the $(i,j)$ entry defined as $\mathbf{K}_{ij} \doteq \kappa(x_i,x_j)$, and $Y \doteq [y_1,\dots,y_T]^\top$, the application of the representer theorem yields $\hat g$ in closed form as follows
\begin{equation}
    \hat g(x) = \sum^T_{j=1} \omega_j \kappa(x, x_j), \, \forall x.
    \label{eqn:reprthm.standard}
\end{equation}
\rev{Thus, considering \eqref{eqn:reprthm.standard} and solving the KRR \eqref{eqn:optprob.init} (see \cite{saunders1998KRR}),} the weights vector $\omega=[\omega_1,\dots,\omega_T]\rev{^{\top}}$ is given by $$\omega=(\mathbf{K}+\gamma \rev{\mathbf{I}_T})^{-1}Y,$$
\rev{with $\mathbf{I}_T$ denoting the $T\times T$ identity matrix.}

The result of this theorem is relevant as it demonstrates that a broad class of learning problems admits solutions that can be expressed as expansions of the training data. Building on this result, the next section explores how the representer theorem extends to the problem of embedding parameterized physical models with data-driven kernels to account for unmodeled dynamics and to estimate interpretable parameters.

\section{Kernel-based model embedding}\label{sec:augkernel}
Let us now consider a {nonlinear map} of the form
\begin{equation}
    y = f(x, \bar \theta) + \Delta(x) + e.
    \label{eqn:system.comp}
\end{equation}
%\textcolor{green}{(it could be better using $\bar \theta$ in this equation, otherwise it may seem that $\theta$ in (4) and in (5) is the same)} 
where $f:\mathcal{X}\times \Theta\to\mathbb{R}$, parametrized in $\bar \theta\in\Theta\subseteq\mathbb R^{n_\theta}$, is a known function derived, e.g., from physical principles, whereas $\Delta:\mathcal{X}\to\mathbb{R}$ is an unknown term representing, e.g., modeling errors, uncertainties, or dynamic perturbations. Let a set of $T$ input-output data $\mathcal{D}=\{(x_1,y_1),\dots,(x_T,y_T)\}$ be given, collected from a realization of \eqref{eqn:system.comp} with true parameters $\bar \theta\in\Theta$. The goal is to find an estimate $\theta^\star$ of $\bar\theta$, and a black-box approximation $\delta:\mathcal{X}\to\mathbb{R}$ of $\Delta(x)$. Such estimate and approximation can be found by solving the following optimization problem:
\begin{equation}
    (\theta^\star, \delta^\star) = \arg \min_{\theta\in\Theta,\delta \in \mathcal{H}} \sum_{t=1}^{T} \left[y_t - \hat y_t\left(\theta,\delta\right)\right]^2 + \gamma\|\delta\|^2_{\mathcal{H}},
    \label{eqn:optprob.theta}
\end{equation}
where $\hat y_t = f(x_t,\theta) + \delta(x_t)$ is the prediction of $y_t$ at time $t$, and the notation $\hat y_t\equiv\hat y_t(\theta,\delta)$ is used to stress its dependencies to $\theta$ and $\delta$. \rev{Note that this formulation is {physics-informed} in the sense that the parametric component~$f(x,\theta)$ embeds known physical relations, while the nonparametric kernel term~$\delta(x)$ accounts only for unmodeled effects or discrepancies with respect to the physical prior. 
This contrasts with standard purely data-driven kernel regression \eqref{eqn:optprob.init}, where no structural knowledge is incorporated.
}

\begin{comment}
{\color{red}
\begin{remark}[On the role of $\gamma$]
{The role of $\gamma$ in~\eqref{eqn:optprob.theta} differs slightly from the one in~\eqref{eqn:optprob.init}. While in~\eqref{eqn:optprob.init} $\gamma$ balances data fit and regularization, in~\eqref{eqn:optprob.theta} it also controls the relative importance assigned to $\delta$ and to the physical model. Specifically, $\gamma$ determines the trade-off between enforcing the structure provided by the physical model and allowing deviations captured by $\delta$. A smaller~$\gamma$ increases the influence of $\delta$, allowing more flexibility in capturing deviations from the physical model, whereas a larger~$\gamma$ enforces stronger adherence to the model structure.}
{For these reasons, a proper tuning of $\gamma$ is necessary to achieve the correct trade-off between physical reliability and model flexibility.} These aspects have been extensively analyzed in~\cite{AutomaticaExtendedVersion}, where the reader can find additional examples and a detailed theoretical discussion.
\end{remark}}
\vspace{.1cm}
\end{comment}
{
\begin{remark}[On the role of $\gamma$]
The hyperparameter~$\gamma$ in~\eqref{eqn:optprob.theta} plays a central role in the proposed framework. 
As in~\eqref{eqn:optprob.init}, it acts as a classical regularization weight balancing data fit and model complexity. 
However, it can also be interpreted as a parameter regulating the relative importance assigned to the physics-based model $f(x,\theta)$ and to the nonparametric correction $\delta(x)$. 
A larger~$\gamma$ enforces stronger adherence to the physical model, promoting smaller and smoother corrections; conversely, a smaller~$\gamma$ increases the influence of~$\delta(x)$, allowing it to capture unmodeled dynamics or finer effects. Conceptually, it serves as an indicator of the relative adequacy of the physical model with respect to the nonparametric correction.
This action introduces a tuning trade-off: %when the physical model already explains most of the system dynamics but residuals contain localized or rapidly varying components (e.g., spikes or high-frequency effects), 
excessively large~$\gamma$ may suppress the flexibility needed for~$\delta(x)$ to represent the residuals. 
Conversely, too small~$\gamma$ may cause~$\delta(x)$ to dominate, effectively ``replacing'' the physical model.
This balance, inherent to any regularization-based method (see, e.g., \cite{tibshirani1996regression,ruppert2004elements}), requires a proper tuning of $\gamma$. For instance, it can be effectively handled through specifically designed selection procedures, such as $k$-fold cross-validation or validation-based tuning, as adopted in this work. 
%Conceptually,~$\gamma$ serves as an indicator of the relative adequacy of the physical model with respect to the nonparametric correction: smaller values are beneficial when the physical structure is incomplete, whereas larger ones are appropriate when the physical model already captures the main system behavior.
%A similar interpretation of the regularization parameter can be found in kernel ridge regression and sparsity-promoting methods such as LASSO~\cite[Chapters~3, 5, and~7]{ruppert2004elements},~\cite{tibshirani1996regression}. 
%Although~\cite{donati2025combining} focuses on predefined basis functions rather than reproducing kernels, the underlying regularization principle is the same:~$\gamma$ controls the trade-off between adherence to prior model structure and correction flexibility.
\end{remark}
}

The optimization problem in ~\eqref{eqn:optprob.theta} aims to estimate the vector of physical parameters associated with the known component of the model $f(x,\theta)$ while simultaneously identifying a function $\delta(x)$ that captures the unmodeled term $\Delta(x)$. This approach embeds available prior physical knowledge $f(x,\theta)$, allows the identification of interpretable parameters $\bar\theta$, and systematically compensates for unmodeled effects $\Delta(x)$, ensuring a more comprehensive and structured representation of the system. 
Assuming that the unknown term $\Delta(x)$ belongs to the RKHS $\mathcal{H}$ associated with the chosen kernel indicates that the solution $\delta^\star$ will admit a kernel representation. Moreover, it also
%
%Assuming that $\Delta$ belongs to a RKHS $\mathcal{H}$ 
implies that $\Delta(x)$ 
%\rev{\sout{is a smooth function that}} 
can be effectively approximated using a finite number of kernel evaluations \rev{parametrized by the observed data points}, as stated in Definition \ref{def:RKHS}. This assumption is common in nonparametric regression \cite{wahba1990spline} and provides a well-posed framework for learning unmodeled dynamics while ensuring regularization and generalization properties. Moreover, although restricting~$\Delta(x)$ to a reproducing space, the RKHSs are flexible enough to approximate a broad class of nonlinear functions \cite{smola2004tutorial}, making this assumption reasonable in many practical systems identification scenarios.

{
%This perspective underscores that 
The primary goal of the identification process is to accurately estimate the physical parameters $\bar\theta$ entering the physical model $f(x,\theta)$. 
The kernel-based representation of $\Delta(x)$ captures and compensates for unmodeled dynamics while preserving the underlying physical structure. This approach ensures that the learned correction term complements the physics-based model rather than overshadowing it.}
The following key result extends the representer theorem to the system identification framework under consideration. \medskip% The proof of the following theorem is reported in Appendix~\ref{apx:repthmII}.

\theorem[Kernel-based model embedding]{\label{thm:repthmII}
Suppose that a nonempty set $\mathcal{X}$, a positive definite real-valued kernel $\kappa$ on $\mathcal X \times \mathcal X$, a dataset $\mathcal{D} = \{(x_1, y_1), \dots, (x_T, y_T)\} \in \mathcal{X}\times \mathbb R$, and a function $f:\mathcal{X}\times \Theta\to\mathbb{R}$, parametrized in $\theta\in\Theta\subseteq\mathbb R^{n_\theta}$ are given. Let us introduce the following functions 
\begin{subequations}
\begin{align}
\Gamma(\theta) &\doteq [f(x_1,\theta), \dots, f(x_T,\theta)]^\top,
\label{eqn:repthm.defs_a}\\
\omega(\theta) &= (\mathbf{K}+\gamma \rev{\mathbf{I}_T})^{-1}(Y-\Gamma(\theta)),
\label{eqn:repthm.defs_b}
\end{align}
\label{eqn:repthm.defs}%
\end{subequations}
where $\mathbf K$ is the kernel matrix associated to $\kappa$ and $\mathcal{D}$, having $\mathbf{K}_{ij}=\kappa(x_i,x_j)$.
Then, Problem~\eqref{eqn:optprob.theta} admits a %unique 
solution $(\theta^\star,\,\delta^\star)$ of the form
\begin{subequations}
\begin{gather}
   %\theta^\star\!\!=\!\arg\min_{\theta\in\Theta} \sum_{t=1}^{T}\!\Bigl(\!y_t\!-\!\hat g(\!z_t,\!\theta)\!-\!\!\sum^T_{j=1} \omega_j(\theta) \kappa(\!z_t,\!z_j)\!\!\Bigr)^2\!\!\!\!\!+\! \gamma\|\Gamma(\theta)\|_\mathcal{H}^2,\\
    \theta^\star\!\!=\!\!\arg\min_{\theta\in\Theta} \sum_{t=1}^{T}\Bigl(\!y_t\!\!-\!\!f(x_t,\theta)\!\!-\!\!\mathbf{K}^\top_t\!\omega(\theta)\!\!\Bigr)^2\!\!\!\!+\!\gamma\rev{\omega(\theta)^{\!\top}\! \mathbf{K}\omega(\theta)},
    \label{eqn:reprthm.ext1}\\
    \delta^\star(x) = \sum^T_{j=1} \omega_j^\star \kappa(x, x_j), \, \omega_j^\star\doteq\omega_j(\theta^\star).
    \label{eqn:reprthm.ext2}
    \end{gather}
    \label{eqn:reprthm.ext}%
\end{subequations}
with $\mathbf{K}^\top_t$ the $t$-th row of $\mathbf{K}$, and $\omega_j(\theta)$ the $j$-th element of~$\omega(\theta)$.
}

%\MM{Do we want to keep proofs in the main text instead of separate appendix? Especially the first one is quite lengthy}
%\CDnote{Add $\omega$ characterization. See, e.g., \cite{care2023kernel}.}

\vspace{.1cm}
\noindent \textit{Proof:} Given \eqref{eqn:optprob.theta} and $\hat y_t = f(x_t,\theta) + \delta(x_t)$, define 
$
J(\theta,\delta) = \sum_{t=1}^{T} \left[y_t - f(x_t,\theta) - \delta(x_t)\right]^2 + \gamma\|\delta\|_{\mathcal{H}},
$
such that \eqref{eqn:optprob.theta} is $(\theta^\star,\delta^\star) = \arg \min_{\theta\in\Theta,\delta \in \mathcal{H}} J(\theta,\delta)$. {Considering that, 
%for a given cost function $J(\alpha,\beta)$ depending on $\alpha\in \mathcal{A}$, $\beta\in \mathcal{B}$, it holds that 
$\min_{\theta\in \Theta,\delta\in\mathcal{H}}J(\theta,\delta)=\min_{\theta\in \Theta} \min_{\delta\in\mathcal{H}}J(\theta,\delta)$,} we have that a minimizer to \eqref{eqn:optprob.theta} must satisfy
\begin{subequations}
\begin{align}
        \delta^\star(\cdot) &= (\arg \min_{\delta\in\mathcal{H}} J(\theta,\delta))|_{\theta=\theta^\star},\\ \theta^\star &= \arg \min_{\theta\in\Theta} p(\theta),
        \label{eqn:thm1.tstar.p}
\end{align}
\end{subequations}
with $p(\theta)\doteq\min_{\delta\in\mathcal{H}} J(\theta,\delta)$. Here, the inner minimization problem 
is solved with respect to $\delta$ and it now represents a standard KRR problem \eqref{eqn:optprob.init}. Indeed, considering $\tilde y_t \doteq y_t - f(x_t,\theta)$, we have
\begin{equation}
\delta^\star (\cdot,\theta) = \arg\min_{\delta\in\mathcal{H}}\sum_{t=1}^{T} \left[\tilde y_t - \delta(x_t)\right]^2 + \gamma\|\delta\|^2_{\mathcal{H}}.
\label{eqn:thm1.dstar}
\end{equation}
By the representer theorem~\cite{scholkopf2001.reprthm}, the optimal solution to \eqref{eqn:thm1.dstar} is 
\begin{equation}
\delta^\star(x,\theta) = \sum^T_{j=1} \omega_j(\theta) \kappa(x, x_j).
\label{eqn:thm1dstarrep}
\end{equation}
\rev{Considering \eqref{eqn:thm1dstarrep} and solving \eqref{eqn:thm1.dstar} as in \cite{saunders1998KRR}} yields the weight vector $\omega(\theta)$ as 
$\omega=(\mathbf{K}+\gamma \rev{\mathbf{I}_T})^{-1}\tilde Y,$ 
being $\tilde Y \doteq [\tilde y_1,\dots,\tilde y_T]^\top$. Thus, \eqref{eqn:repthm.defs_b} is obtained given \eqref{eqn:repthm.defs_a} and substituting $\tilde y_t \doteq y_t - f(x_t,\theta)$ in $\tilde Y$.
Moreover, considering the function $p(\theta)$ in \eqref{eqn:thm1.tstar.p}, we have $p(\theta)=\min_{\delta\in\mathcal{H}}J(\theta,\delta)=J(\theta,\delta^\star(\cdot,\theta))$, which, substituting \eqref{eqn:thm1dstarrep}, simplifies to
\begin{equation}
    p(\theta) = \textstyle\sum_{t=1}^{T}(y_t\!-\!f(x_t,\theta)\!-\!\mathbf{K}^\top_t\omega(\theta))^2\!+\!\gamma
    \rev{\omega(\theta)^\top \mathbf{K}\omega(\theta)},
    %\|\sqrt{\mathbf{K}}\omega(\theta)\|_2^2,
    \label{eqn:thm1pstar}
\end{equation}
noting that 
$$
\begin{aligned}
\|\delta\|_\mathcal{H}^2&=\langle\textstyle\sum^T_{i=1} \omega_i(\theta) \kappa(x, x_i),\textstyle\sum^T_{j=1} \omega_j(\theta) \kappa(x, x_j)\rangle_\mathcal{H}\\
&=\textstyle\sum^T_{i=1}\textstyle\sum^T_{j=1} \omega_i(\theta) \omega_j(\theta)\langle \kappa(x, x_i), \kappa(x, x_j)\rangle_\mathcal{H}\\
&=\textstyle\sum^T_{i=1}\textstyle\sum^T_{j=1} \omega_i(\theta) \omega_j(\theta) \kappa(x_i, x_j)\\ &= \omega(\theta)^\top \mathbf{K}\omega(\theta),
%\rev{\sout{ = \|\sqrt{\mathbf{K}}\omega(\theta)\|_2^2,}}
\end{aligned}
$$
from linearity of the inner product and the reproducing property in Definition~\ref{def:RKHS}.b.   %\rev{\sout{Here, $\sqrt{\mathbf{K}}$ denotes the principal square root of $\mathbf{K}=\sqrt{\mathbf{K}}\sqrt{\mathbf{K}}=\sqrt{\mathbf{K}}^\top\sqrt{\mathbf{K}}$, noting that $\mathbf{K}$ is guaranteed to be symmetric and at least positive semi-definite by Definition \ref{def:pdkernel}.}}
Thus, we obtain \eqref{eqn:reprthm.ext} by substituting the solution to \eqref{eqn:thm1.tstar.p}, with $p(\theta)$ given by \eqref{eqn:thm1pstar}, into \eqref{eqn:thm1dstarrep}, which concludes the proof. \hfill $\square$ \smallskip

Theorem~\ref{thm:repthmII} establishes that the optimal solution to the estimation problem \eqref{eqn:optprob.theta} can be formulated using kernel-based functions. In particular, the unmodeled component $\Delta(x)$ is approximated by $\delta(x)$, defined as a linear combination of kernel evaluations \rev{parametrized by the observed data points}, thus leading to the following 
%\rev{\sout{optimal}} 
predictive model, \rev{representing the optimal solution to Problem~\eqref{eqn:optprob.theta}:}
\begin{equation*}
    \hat y = f(x, \theta^\star) + \delta^\star(x) = f(x, \theta^\star)+\textstyle\sum^T_{j=1} \omega_j^\star \kappa(x, x_j).
\end{equation*}
Theorem~\ref{thm:repthmII} is particularly relevant as, in contrast to standard KRR, it enables the explicit incorporation of prior physical knowledge alongside the adaptability of kernel methods, {avoiding the use of heuristically chosen basis functions}.

\rev{
\begin{remark}[On hyperparameter tuning]
Clearly, kernel methods still involve hyperparameters (e.g., the kernel bandwidth $\sigma$ in Gaussian and Laplacian kernels), which are typically tuned heuristically or via validation. This issue, however, is not unique to kernels: dictionary-based methods also require hyperparameter choices, such as the regularization weights that promote sparsity, as well as parameters embedded in the basis functions themselves \cite{AutomaticaExtendedVersion}, concluding that some level of hyperparameter tuning is unavoidable in both approaches. Nevertheless, kernel-based models generally rely on a smaller number of hyperparameters, which simplifies the identification procedure compared to dictionary-based alternatives.
\end{remark}
}

%\MM{\textbf{I believe this paragraph would effectively strengthen the contribution in the introduction: }Unlike traditional approaches that rely on predefined dictionaries of basis functions, often selected heuristically and without a clear guiding principle, the proposed kernel-based formulation inherently adapts to the underlying data structure. Thus, by leveraging the representer theorem, this framework ensures a data-driven and regularized functional approximation, eliminating the need for the manual selection of the basis functions while preserving both interpretability and accuracy thanks to the underlying physical model.} 

\subsection{{Affine-in-parameters models}}
%Following Theorem \ref{thm:repthmII}, 
A relevant special case arises when the physics-based function $f(x, \theta)$ is {affine} in $\theta$. In this setting, the map \eqref{eqn:system.comp} becomes
%\GC{[I think affine, rather than linear... :}
{\begin{equation}
y = f_0(x) + f(x)^\top \bar{\theta} + \Delta(x) + e,
\label{eqn:system.comp.lin}%
\end{equation}}%
%Note that we cannot simply say that $\bar\theta$ contains a fixed "1" to account for the constant term, because in formula (14) you are computing $\theta$ and cannot impose such structure. The derivations below should be modified a bit by using $y-f_0(x)$ instead of $y$...]
%\begin{equation}
%y = f(x)^\top \bar{\theta} + \Delta(x) + e,
%\label{eqn:system.comp.lin}
%\end{equation}
where {$f_0:\mathcal{X}\to\mathbb{R}$}, $f:\mathcal{X}\to\mathbb{R}^{n_\theta}$, $\bar \theta\in\Theta\in\mathbb R^{n_\theta}$.
Thus, the optimization problem \eqref{eqn:reprthm.ext1} simplifies significantly, leading to a convex optimization problem with a closed-form solution, as formalized in the following theorem.
\medskip
\begin{theorem}[Closed-form solution of \eqref{eqn:reprthm.ext}]\label{thm:retpthmII.lin}
Consider the same setup of Theorem \ref{thm:repthmII}. \rev{Define $F(x) \doteq [f(x_1), \dots, f(x_T)]^\top \in \mathbb R^{T,n_\theta}$ and $Y_0 \doteq [y_1-f_0(x_1), \dots, y_T-f_0(x_T)]^\top \in \mathbb R^T$. Assume $F(x)$ is full column rank.} If the system model in \eqref{eqn:system.comp} is {affine} in $\theta$, as in \eqref{eqn:system.comp.lin}, then the solution of \eqref{eqn:reprthm.ext1} is given by
\begin{equation}
\theta^\star = \left(F(x)^\top \rev{\Psi} F(x)\right)^{-1}F(x)^\top \rev{\Psi} {Y_0},
\label{eqn:thmlincf}
\end{equation}
with
\begin{comment}
\begin{equation}
    \Psi = \left[\begin{array}{c}
\rev{\mathbf{I}_T}-\mathbf{K}(\mathbf{K}+\gamma \rev{\mathbf{I}_T})^{-1}\\
{\gamma}\sqrt{\mathbf{K}}(\mathbf{K}+\gamma \rev{\mathbf{I}_T})^{-1}
\end{array}\right],
\label{eqn:Psidef}
\end{equation}
\end{comment}
\begin{comment}
\begin{equation}
    \rev{\begin{array}{c}
    \Psi \doteq \Psi_1^\top\Psi_1 + \Psi_2,\\
\Psi_1\doteq{\mathbf{I}_T}-\mathbf{K}(\mathbf{K}+\gamma {\mathbf{I}_T})^{-1},\\
\Psi_2\doteq\gamma[(\mathbf{K}+\gamma \mathbf{I}_T)^{-1}]^\top \mathbf{K} [(\mathbf{K}+\gamma \mathbf{I}_T)^{-1}],
\end{array}}
\label{eqn:Psidef}
\end{equation}
\end{comment}
\begin{equation}
    \rev{
    \Psi \doteq (\mathbf{K}+\gamma {\mathbf{I}_T})^{-1}},
\label{eqn:Psidef}
\end{equation}
\begin{comment}
\begin{equation}
    \rev{\Psi \doteq \!\!\!\begin{array}{l}
[{\mathbf{I}_T}-\mathbf{K}(\mathbf{K}+\gamma {\mathbf{I}_T})^{-1}]^\top[{\mathbf{I}_T}-\mathbf{K}(\mathbf{K}+\gamma {\mathbf{I}_T})^{-1}]\\
+ \gamma[(\mathbf{K}+\gamma \mathbf{I}_T)^{-1}]^\top \mathbf{K} [(\mathbf{K}+\gamma \mathbf{I}_T)^{-1}]
\end{array}\!\!\!,}
\label{eqn:Psidef}
\end{equation}
\end{comment}
where $\mathbf K$ is the kernel matrix associated to $\kappa$ and $\mathcal{D}$, \rev{and $\gamma$ is the weight controlling the regularization of $\delta(\cdot)$ \eqref{eqn:reprthm.ext2}}.
\end{theorem}

\noindent \textit{Proof:} Considering \eqref{eqn:reprthm.ext1} applied to \eqref{eqn:system.comp.lin}, we obtain
\begin{equation}
\begin{aligned}
\theta^\star=\arg\min_{\theta\in\Theta} &\sum_{t=1}^{T}\Bigl(y_t{-f_0(x_t)}-f(x_t)^\top\theta-\mathbf{K}^\top_t\omega(\theta)\Bigr)^2\\&+\gamma\rev{\omega(\theta)^{\top}\mathbf{K}\omega(\theta)}.
\end{aligned}
\end{equation}
{Consider the centered output $y_t-f_0(x_t)$.} 
%\rev{To simplify notation, let $\mathbf{M}\doteq(\mathbf{K}+\gamma \rev{\mathbf{I}_T})^{-1}$.} 
Rewriting the summation and substituting \eqref{eqn:repthm.defs} and \eqref{eqn:Psidef}, we obtain
\begin{equation}
\begin{aligned}
\theta^\star=\arg\min_{\theta\in\Theta} &\|{Y_0}-F(x)\theta-\mathbf{K}\,\rev{\Psi}({Y_0}-F(x)\theta)\|_2^2\\
%&+\gamma\|\sqrt{\mathbf{K}}(\mathbf{K}+\gamma \rev{\mathbf{I}_T})^{-1}({Y_0}-F(x)\theta)\|_2^2,\\
&+\rev{\gamma({Y_0}-F(x)\theta)^\top\Psi^\top \mathbf{K}\, \Psi({Y_0}-F(x)\theta)}
\end{aligned}
\label{eqn:linproof1}
\end{equation}
{where, 
%considering the centered output $y_t-f_0(x_t)$,
we used the fact that \eqref{eqn:repthm.defs_a} corresponds to $\Gamma(\theta)=F(x)\theta$, and, according to \eqref{eqn:repthm.defs_b} and \eqref{eqn:Psidef}, $\omega(\theta) = (\mathbf{K}+\gamma \rev{\mathbf{I}_T})^{-1}(Y_0-F(x)\theta)\rev{=\Psi(Y_0-F(x)\theta)}$. }
To simplify this expression further, consider 
\begin{subequations}
\begin{gather}
\Psi_1 \doteq \rev{\mathbf{I}_T}-\mathbf{K}\,\rev{\Psi}, \label{eqn:Psi1def}\\
%\rev{\sout{$\Psi_2 \doteq \gamma \sqrt{\mathbf{K}}(\mathbf{K}+\gamma \rev{\mathbf{I}_T})^{-1}$}}
\rev{\Psi_2 \doteq \gamma\Psi^\top \mathbf{K} \,\Psi}.\label{eqn:Psi2def}
\end{gather}
\end{subequations}
% from \eqref{eqn:Psidef}. 
Substituting these definitions into \eqref{eqn:linproof1}, we write 
\begin{equation}
\begin{aligned}
\theta^\star=\arg\min_{\theta\in\Theta} &\|\Psi_1 ({Y_0}-F(x)\theta)\|_2^2 \\&\rev{+ ({Y_0}-F(x)\theta)^\top\Psi_2({Y_0}-F(x)\theta)}.
\end{aligned}
\label{eqn:prooflin2}
\end{equation}
{Problem \eqref{eqn:prooflin2} can be recognized as a standard \rev{weighted} least-squares problem, since the objective is quadratic in ${Y_0}-F(x)\theta$, and can be written compactly as
$$ \rev{{\bigl({Y_0}-F(x)\theta\bigr)^\top\left(\Psi_1^\top\Psi_1+\Psi_2 \right) \bigl({Y_0}-F(x)\theta\bigr)}.} %= {\Bigl\|\Psi \bigl({Y_0}-F(x)\theta\bigr)\Bigr\|_2^2}
$$
%$\|a\|^2_2+\|b\|^2_2 = \|[a^\top, b^\top]^\top\|^2_2$.
%of the form $\theta^\star = \arg\min_{\theta \in \Theta}\|A\theta-b\|_2^2+\|C\theta-d\|_2^2$ \cite{golub1999tikhonov}, 
%\rev{Being $\Psi\doteq\Psi_1^\top\Psi_1 + \Psi_2$ \eqref{eqn:Psidef}, 
\rev{Therefore, the optimal solution has the closed form 
\begin{equation}
\theta^\star \!=\! \left[F(x)^\top {(\Psi_1^\top\Psi_1 + \Psi_2)} F(x)\right]^{-1}F(x)^\top {(\Psi_1^\top\Psi_1 + \Psi_2)} {Y_0}.
\label{eqn:Thm2long}
\end{equation}
%\eqref{eqn:thmlincf}.
%where expanding $\mathbf{M}$ in the definitions of $\Psi_1$, $\Psi_2$ yields \eqref{eqn:Psidef}.
\rev{To further simplify this expression, recall from the definition of $\Psi$ in \eqref{eqn:Psidef} that $(\mathbf{K}+\gamma \mathbf{I}_T)\Psi=\mathbf{I}_T.$ Rearranging, this implies
$$\gamma\Psi=\mathbf{I}_T-\mathbf{K}\Psi.$$
Now, using this relation in \eqref{eqn:Psi1def}, we obtain
$$
\Psi_1=\mathbf{I}_T - \mathbf{K}\Psi=\gamma \Psi.
$$}
Substituting into $\Psi_1^\top\Psi_1 + \Psi_2$ yields
$$
\begin{aligned}
    \Psi_1^\top\Psi_1 + \Psi_2 &= \gamma^2\Psi^\top\Psi + \gamma\Psi^\top \mathbf{K}\Psi \\ &= \gamma\Psi^\top(\gamma \mathbf{I}_T+\mathbf{K})\Psi,
\end{aligned}
$$
factoring out $\gamma\Psi^\top$ and $\Psi$. Hence, being $(\mathbf{K}+\gamma \mathbf{I}_T)\Psi=\mathbf{I}_T$ from \eqref{eqn:Psidef}, and $\Psi$ symmetric, we conclude
\begin{equation}
\Psi_1^\top\Psi_1 + \Psi_2=\gamma\Psi.
\label{eqn:laststep}%
\end{equation}%
Substituting \eqref{eqn:laststep} into \eqref{eqn:Thm2long} directly yields \eqref{eqn:thmlincf},
with the scalar factor $\gamma$ canceling out as it appears both inside the inverse and outside. This
concludes the proof.}}
%for which the optimal solution is given in closed form by} 
\begin{comment}
$$\theta^\star \doteq \left(\left[\begin{array}{c}
     A\\
     C 
\end{array}\right]^\top\left[\begin{array}{c}
     A\\
     C 
\end{array}\right]\right)^{-1}\left[\begin{array}{c}
     A\\
     C 
\end{array}\right]^\top\left[\begin{array}{c}
     b\\
     d 
\end{array}\right].$$
\end{comment}
%Thus, 
%\CD{\eqref{eqn:thmlincf}, 
%follows considering $A = \Psi_1F(x)$, $b = \Psi_1Y$, $C = \Psi_2F(x)$, $d = \Psi_2Y$, and $\Psi = [\Psi_1^\top, \Psi_2^\top]^\top$, 
%concluding the proof.}
%\GC{[Never heard of Tikhonov regularization in this sense... Just say it is plain LS, since $\|a\|^2+\|b\|^2 = \|(a, b)\|^2$ and skip the above block formula...]}
\hfill $\square$

\rev{
\begin{remark}[On matrix invertibility and system identifiability]
It is worth noting that $\Psi$ is always positive definite, being defined as the inverse of the matrix 
$(\mathbf{K} + \gamma \mathbf{I}_T)$. Indeed, $\mathbf{K}$ is guaranteed to be symmetric and at least positive semidefinite 
by Definition~\ref{def:pdkernel}, and $\gamma \mathbf{I}_T \succ 0$ for any $\gamma > 0$. 
Consequently, the only requirement for the invertibility of $F(x)^\top \Psi F(x)$ is that 
$F(x)$ has full column rank, as assumed in Theorem~\ref{thm:retpthmII.lin}. 
The full column rank condition requires that $T \geq n_\theta$ (i.e., at least as many data points as parameters) and that the regressor vectors ${f}({x}_t)$ are linearly independent. This is equivalent to requiring that the input signal is {persistently exciting}.
If $F(x)$ is not full column rank, the solution to \eqref{eqn:reprthm.ext1} is not unique, 
reflecting an identifiability issue due to insufficient excitation in the input signal. 
In this case, one can select the minimum-norm solution, obtained by replacing the inverse 
with the Moore--Penrose pseudoinverse, i.e.,
$$
\theta^\star = \bigl(F(x)^\top \Psi F(x)\bigr)^\dagger F(x)^\top \Psi Y_0.
$$
\end{remark}
}

{This result is particularly significant as it shows that, when the model is {affine} in $\theta$, the optimization problem \eqref{eqn:reprthm.ext1} becomes convex, ensuring a unique and efficiently computable solution. }
\rev{Indeed, the computation of the closed-form solution in \eqref{eqn:thmlincf} requires only standard matrix inversions. On the other hand, in the non-affine case, computing the solution to \eqref{eqn:reprthm.ext1} requires iteratively computing the gradient and evaluating the kernel, thus leading to higher computational cost.
}

Importantly, the structure in \eqref{eqn:system.comp.lin} is quite general, as it does not impose linearity with respect to the input $x$, but only in the parameters. Notably, many nonlinear  (with respect to their inputs) systems can still be expressed in this form, making the framework and Theorem \ref{thm:retpthmII.lin} broadly applicable.
%Furthermore, many nonlinear system models can be reformulated in a structure that is linear in parameters (see, e.g., \cite{} ), making this theorem broadly applicable in practical system identification scenarios. 
On the other hand, {when affinity in the parameters does not hold, we can still tackle problem \eqref{eqn:reprthm.ext1} 
by means of nonlinear programming methods, such as gradient-based techniques, Gauss-Newton-type algorithms, or similar iterative approaches, acknowledging however that due to the potential non-convexity of the problem, the obtained solution may be local.
}
%In the next section, we extend the kernel-based framework to state-space systems, enabling its application to a broader class of dynamical models.

\section{Application to state-space systems}\label{sec:sskernel}
{In the previous section, we considered a static input-output identification setting, where the goal was to estimate physical parameters $\bar\theta$ exploiting physical priors $f(x,\theta)$ and approximating an unknown function $\Delta(x)$ based on measured data pairs $(x_t, y_t)$. 
%However, many physical systems are better described by state-space models, which explicitly capture system dynamics over time \cite{care2023kernel}. Furthermore, this formulation unifies various prediction models, including nonlinear output error, ARMAX, and ARX models \cite{ribeiro2020smoothness}, which makes it particularly valuable for system identification. Additionally, many controller and observer design approaches are based on a state-space representation. Unlike static regression models, a state-space formulation accounts for the evolution of hidden states, requiring estimating both system parameters and unmeasured state trajectories.
In this section, we extend the kernel-based framework to dynamic settings characterized by not-fully measured states by introducing a state-smoothing approach to effectively address the associated challenges.
}
%{In many cases, a state-space representation provides a more effective framework for describing physical phenomena \cite{care2023kernel}. \textcolor{red}{Moreover, many controller and observer design approaches are based on a state-space representation.} %\textcolor{green}{(a detail: 'state-space' is more suitable for linear systems; 'state-equation' is perhaps more appropriate; 'state-space' can be fine ...)}
%This formulation unifies various prediction models, including nonlinear output error, ARMAX, and ARX models \cite{ribeiro2020smoothness}, making it particularly valuable for system identification. 

{We consider a discrete-time system of the form}
%\textcolor{red}{the system in~\eqref{eqn:system.comp}} can be 
%
\begin{equation}
    x_{t+1} = f(x_t, u_t, \bar\theta) + \Delta(x_t,u_t) + e_t,
    \label{eqn:system.ssform}
\end{equation}
where $x_t\in\mathbb{R}^n$ denotes the state at time $t$, $u_t \in\mathbb R^{n_u}$ is the external, measured input, and $e_t\in\mathbb{R}^n$ represents the process noise. {The physics-based function $f(x_t,u_t,\theta)$ and the unmodeled dynamics $\Delta(x_t,u_t)$ are now vector-valued, each comprising $n$ components, i.e., $f_i(x_t,u_t,\bar\theta_i)$ and $\Delta_i(x_t,u_t)$, $i = 1, \dots, n$. If {all state variables are directly measurable,} each parameter vector $\bar\theta_i$ and function $\Delta_i$ can be estimated using Theorem~\ref{thm:repthmII} directly. In this case, the state $x_{t}$ serves both as the input -- along with the measured input $u_{t}$ -- and as the measured output ($y_{t} = x_{t}$),
allowing for a direct application of Theorem~\ref{thm:repthmII}. However, this approach becomes infeasible when certain state components are not directly measurable. In such cases, the system %\eqref{eqn:system.comp} 
\eqref{eqn:system.ssform}
is extended to incorporate also the output equation, i.e.,}
\begin{equation}
    \begin{aligned}
    x_{t+1} &= f(x_t, u_t, \bar\theta) + \Delta(x_t,u_t) + e_t,\\
    y_t &= g(x_t,u_t, \bar\theta) + w_t,
    \end{aligned}
    \label{eqn:system.sswoform}
\end{equation}
where $w_t$ represents the measurement noise, and the state $x_t$ is not directly accessible. {However, this framework adds a further layer of complexity to the estimation procedure, which can be addressed by adopting two principal approaches: (i) multi-step identification methods, or (ii) prior states estimation.}
%Although multi-step identification provides a natural way to estimate latent states through the available physics, it often compromises the convexity of the associated cost functions, leading to challenging optimization problems due to the non-linear dependencies between parameters. Furthermore, extending the kernel-based framework to multi-step settings introduces additional complexity as the recursive dependence of $\delta$ within $f$, and thus $\theta$, prevents a straightforward definition of $\delta$ through the application of Theorem~\ref{thm:repthmII}.
{As anticipated in the introduction, the optimization becomes challenging in multi-step identification due to significant nonlinear dependencies among the parameters. Moreover, the recursive dependence of $\delta(x_t,u_t)$ within $f(x_t,u_t,\theta)$, and consequently~$\theta$, prevents a straightforward application of Theorem~\ref{thm:repthmII}.
To circumvent this issue, in the following we focus on the second class of approaches, adopting nonlinear state smoothing techniques.}
%to enhance the kernel-based estimation framework.}

\subsection{Nonlinear state reconstruction}
We consider the system described by \eqref{eqn:system.sswoform}, where the state variable $x_t\in\mathcal{X}$ evolves according to known physics-based functions $f:\mathcal{X}\times\mathbb R^{n_u}\times\Theta\to\mathcal{X}$, and $g:\mathcal{X}\times\mathbb R^{n_u}\times\Theta\to\mathbb R$, and an unknown term $\Delta:\mathcal{X}\times\mathbb R^{n_u}\to\mathcal{X}$, {related to unmodeled dynamics}. The goal of nonlinear state smoothing is to estimate a state trajectory ${x}_{0:T-1}\doteq\{ x_0,..., x_{T-1}\}$ from a given dataset of measurements $\mathcal{D}=\{(u_0,y_0),\dots,(u_{T-1},y_{T-1})\}\cup\{y_T\}$ and a nominal nonlinear model \cite{sarkka2008unscented}.

%First, we consider the known elements of \eqref{eqn:system.sswoform}, i.e.,
{First, let us consider the known components of \eqref{eqn:system.sswoform}, which define the nominal model, i.e.,}
\begin{equation}
    \begin{aligned}
    x_{t+1} &= f(x_t, u_t, \theta_0) + e_t,\\
    y_t &= g(x_t,u_t, \theta_0) + w_t,
    \end{aligned}
    \label{eqn:system.sswoform.init}
\end{equation}
%as the nominal model 
where $\theta_0$ represents the initial parameter estimate used for the state smoothing. This can correspond, for example, to an initial guess or to the central point in the parameter space $\Theta$, which will be refined during the identification process. 
\begin{comment}
\CD{
\begin{remark}[model uncertainty in state estimation]
    The use of an approximated model for predictions in the filtering process, combined with the fact that $\theta_0$ represents only a nominal initial guess, inherently introduces uncertainty in the estimated state trajectories. This uncertainty can be addressed in several ways. A common approach is to incorporate the modeling error into the process and measurement noise terms, $w_t$ and $e_t$, by appropriately tuning their covariance matrices within the filter algorithm to mitigate the effects of model discrepancies. An alternative strategy involves an iterative refinement approach, where the initial model used in the filtering step is progressively updated, and the noise covariance matrices are adjusted after each identification cycle, until the parameter estimates converge. However, we remark that a perfect model is not required in this stage; the primary objective is to obtain sufficiently accurate state estimates to support the subsequent identification process. Theoretical studies on the bounds of the prediction errors in the smoothing step are currently under investigation and will be the subject of future work.
\end{remark}
}
\end{comment}
%
Moreover, without loss of generality, we assume that an initial estimate of the initial condition, denoted as $\hat x_0$, is available. This estimate can be seamlessly incorporated into the identification problem alongside $\theta$ if needed (see, e.g., \cite{AutomaticaExtendedVersion}). 

To perform the nonlinear state reconstruction, in this work, we employ a nonlinear state smoothing based on a two-step strategy. First, we apply a \textit{forward filtering}, based on an unscented Kalman filter, {for state estimation}. Then, we employ a \textit{backward smoothing} for refining the state estimates. However, we note that any state reconstruction approach can be employed in this last step. 

{Therefore, we begin by imposing standard conditions ensuring that the state can be estimated and the physical parameters can be identified from the available measurements.
\begin{assumption}[observability and identifiability]
    The system state $x$ is observable along the trajectory induced by the applied input, and the physical parameters $\theta$ are identifiable~\cite{bellman1970structural}, provided that the input is persistently exciting over the observation window.
\end{assumption}}
Next, we detail the proposed two-step approach as applied to the considered state-space framework.\vspace{.2cm}

\noindent
\textit{1. Forward filtering}: The UKF aims to approximate the posterior state distribution using a set of sigma points, which are propagated through the nominal nonlinear system dynamics.
% \cite{wan2000unscented}. %\CD{Here, we assume that the process noise $e_t$ and measurement noise $w_t$ can be characterized as zero-mean Gaussian noise terms with covariances $P_e$ and $P_w$, respectively:$e_t \sim \mathcal{N}(0, P_e)$, $w_t \sim \mathcal{N}(0, P_w)$}. \CD{The covariance matrix $P_t$ represents the uncertainty associated with the state estimate $\hat x_t$. It is initialized as $P_0$ and updated recursively through the filtering process.} 
{Here, we assume that the process and measurement noise, $e_t$ and $w_t$, can be characterized as zero-mean Gaussian with covariances $P_e$ and $P_w$, respectively, i.e., $e_t \sim \mathcal{N}(0, P_e)$, $w_t \sim \mathcal{N}(0, P_w)$. The state estimate uncertainty, represented by $P_t$, is initialized as $P_0$ and updated recursively.}
All common variants of the UKF for discrete-time systems adhere to the same prediction–correction structure, though they may differ in specific formulations and weight definitions. %In some cases, for instance, the state is augmented to incorporate process and measurement noise. 
The reader is referred to, e.g., \cite{wan2000unscented} 
%,julier2004unscented,menegaz2015systematization} 
for %additional 
details on the UKF and its implementation.
%\MM{In my opinion, if we need to save some space I would remove this last paragraph or shrink it a lot to simply ``\textit{The reader is referred to, e.g., \cite{wan2000unscented} for additional details on the UKF and its implementation.}''}

%At the end of the forward filtering step, we obtain the filtered state sequence $\hat x_{1:T}\doteq\{\hat x_1,\dots,\hat x_T\}$ with the associated covariance matrices $P_{1:T}\doteq\{P_1,\dots,P_T\}$, which will be used as the input to the smoothing process, described in the next section.

\vspace{.2cm}
{At the end of the forward filtering step, we obtain the filtered state sequence $\hat x_{1:T}\doteq\{\hat x_1,\dots,\hat x_T\}$ with the associated covariance matrices $P_{1:T}\doteq\{P_1,\dots,P_T\}$. These estimates serve as the input for the subsequent smoothing process, which is described next.}
\vspace{.2cm}

\noindent
\textit{2. Backward smoothing}: 
The
%backward smoothing phase is based on the 
unscented Rauch–Tung–Striebel smoother  \cite{sarkka2008unscented,akhtar2023computationally} aims to obtain the final state estimates. Given the filtered estimates, sigma points are generated and propagated through the model to compute predicted means, covariances, and cross-covariances \cite{sarkka2008unscented}. The smoother gain matrix is then used to iteratively refine the state and covariance estimates, running backward from time $t = T-1$ to $t = 0$, yielding the final smoothed state sequence $\hat x^s_{0:T-1}\doteq\{\hat x^s_0,\dots,\hat x^s_{T-1}\}$ and covariances $P^s_{0:T-1}\doteq\{P^s_0,\dots,P^s_{T-1}\}$.

\begin{comment}
\CD{
\begin{remark}[model uncertainty in state estimation]
    Clearly, using an approximated model and a nominal $\theta_0$ introduces uncertainty in state estimates. This can be managed, for instance, by incorporating the modeling error into the process and measurement noise terms, $w_t$ and $e_t$, appropriately tuning their covariance matrices within the filter and smoother algorithms. Alternatively, an iterative refinement approach can be adopted: the initial model used in the filtering step is progressively updated, adjusting accordingly the noise covariance matrices after each identification cycle, until convergence. Future work will explore this and error bounds in the smoothing step.
\end{remark}
}
\GC{[Ttralascerei questo remark. Magari si può poi dire qualcosa nelle conclusioni.]}
\end{comment}

\subsection{State-space reformulation}
Once the sequence of unmeasured states has been reconstructed through state smoothing, i.e., $\hat x^s_{0:T-1}$, the problem effectively reduces to the one in \eqref{eqn:optprob.theta}, to which Theorem~\ref{thm:repthmII} is directly applicable. This can be done, for instance, by defining $z_t \doteq [\hat x^s_{t-1}, u_{t-1}, u_t]$ as the \emph{new} input variable, and computing the predictions at time $t$, i.e., $\hat y_t(\theta,\delta)$ in \eqref{eqn:optprob.theta}, using the following prediction model
\begin{equation}
    \hat y_t = \xi(z_t,\theta) + \delta(z_t),
    \label{eqn:predmodel.new}
\end{equation}
with $\xi(z_t)\doteq g\big(f(\hat x^s_{t-1},u_{t-1},\theta),u_t,\theta\big)$.
In this formulation,~$\delta(z_t)$ captures the discrepancies arising from unknown or unmodeled components in $f(x_t,u_t,\theta)$ \eqref{eqn:system.sswoform}, which influence the system evolution and are subsequently mapped to the output space by $g(x_t,u_t,\theta)$. 

%\textcolor{green}{We want to remark that despite there exist alternative models, the selected formulation fully leverages the available physical priors, incorporating both~$f$~and~$g$. Correspondingly, the proposed methodology allows for both estimating the physical parameters $\theta$ and approximating the unknown component~$\Delta$ using the kernel-based approach, effectively leveraging the available dataset and \emph{all} the available physical information. Hence, the proposed framework ensures the principled integration of prior knowledge with data-driven modeling, enhancing both interpretability and accuracy. 
%{Furthermore, the inclusion of the nonlinear state smoother allows us to preserve the well-behaved properties of single-step identification while inherently capturing multi-step dependencies. Indeed, the multi-step propagation of the system dynamics is implicitly handled within the smoothing process, which provides state estimates that serve as inputs to the identification problem. As a result, the model benefits from improved accuracy in both single-step and multi-step tasks, without requiring explicit multi-step optimization, and providing accuracy in simulation-based applications.}}
%\GC{[Can't we cut a couple of sentences from the green part above?]}
%\MM{I would move this part in red at the end of this section (or right before the last sentence introducing the Algorithm) and possibly stated as a remark.}
\vspace{.1cm}
{\begin{remark}[estimation without future data]
When employing the identified model for simulation, where both the learned kernel component and the identified parameters are used, future data is not always available to apply the smoother. In this case, the natural solution is to rely solely on the filtering step to provide state estimates up to time $t$. Beyond this point, the nominal model is used in open-loop to propagate the unmeasured states, which are then fed into the kernel-extended model to generate output predictions. 
\end{remark}}

\vspace{.1cm}
{We note that, although alternative models exist, the selected formulation fully leverages the available physical priors by incorporating both $f(x_t,u_t,\theta)$ and $g(x_t,u_t,\theta)$. The proposed methodology enables the estimation of physical parameters $\theta$ and the approximation of the unknown component $\Delta(x_t,u_t)$ via a kernel-based approach, effectively embedding prior knowledge with data. This principled combination enhances both interpretability and accuracy. Additionally, the nonlinear state smoother preserves the well-behaved properties of single-step identification while implicitly capturing multi-step dependencies. This leads to improved accuracy in both single- and multi-step settings, without requiring explicit multi-step optimization.}

The overall procedure for implementing the proposed identification approach in a state-space setting is sketched in Algorithm~\ref{alg:kernel_state_space}. 
\begin{algorithm}[!tb]
\caption{{Kernel-based physics-informed identification}}
\label{alg:kernel_state_space}
\begin{algorithmic}[1]
    \STATE \textbf{Input:} Dataset $\mathcal{D}=\{(u_0,y_0),\dots,(u_{T-1},y_{T-1})\}\cup\{y_T\}$, $f$, $g$, $\theta_0$, $\hat{x}_0$, $P_0$, $P_e$, $P_w$,
    %$a$, $w^m$, $w^c$, 
    $\kappa$, \rev{$\gamma$}.
    %\STATE \textbf{Step 1: UKF}
    \STATE $\hat x_{1:T}, P_{1:T} \gets$ UKF \cite{wan2000unscented}.
    \hspace{\algorithmicindent} 
    %\STATE Obtain filtered states $ \hat{x}_{1:T}$ and covariances $P_{1:T}$.
    %\STATE \textbf{Step 2: URTSS}\\
    \STATE $\hat x^s_{0:T-1}, P^s_{0:T-1} \gets $ Rauch–Tung–Striebel smoother  \cite{sarkka2008unscented}.
    %\STATE Obtain smoothed states $ \hat{x}^s_{1:T}$ and covariances $P^s_{1:T}$.
    %\STATE \textbf{Step 3: Solve the kernel-based optimization problem}
    \STATE Define new input $ z_t \doteq [\hat{x}^s_{t-1}, u_{t-1}, u_t] $.
    \STATE Define a new prediction model with input $z_t$, as in \eqref{eqn:predmodel.new}.
    \STATE Apply Theorem \ref{thm:repthmII}, and solve \eqref{eqn:reprthm.ext1} with any preferred optimization method, or with Theorem \ref{thm:retpthmII.lin}, if affine in $\theta$.
    \STATE \textbf{Output:} Estimated parameters $ \theta^\star $ and function $ \delta^\star $.
\end{algorithmic}
\end{algorithm}

\begin{comment}
Once the smoothed state trajectory $\hat x^s_{0:T-1}$ is obtained, it can be used alongside the measured input $u_0,\dots,u_{T-1}$, as input to an estimation model of the form 
\begin{equation}
    \hat y_t = g(f(\hat x^s_{t-1},u_{t-1},\theta),u_t,\theta) + \delta(\hat x^s_{t-1},u_{t-1},u_t),
\end{equation}
where $\delta$ captures discrepancies arising from unknown or unmodeled components in the state transition function $f$ in \eqref{eqn:system.sswoform}, which influence the system's evolution and are subsequently mapped to the output space by $g$.
\end{comment}

\section{Numerical example}\label{sec:results}
In this section, we illustrate  the effectiveness of the proposed identification method on {an academic example} and a cascade tank system (CTS) benchmark \cite{schoukens2016cascaded}.
% Rem. Cit.: ,schoukens2017three}, 
\subsection{Academic example}
We consider a regression problem where the goal is to estimate the parameters $\bar \theta$ of a linear-in-parameter model, using the kernel-based approach and comparing it to the solution obtained when no kernel embedding is employed, solved with ordinary least-squares methods, \rev{the discrepancy modeling approach, and the solution to the standard kernel ridge regression problem (i.e., without embedding prior physical knowledge)}. 

%%%%%%%%%%%%%%%%%%%%%%%%%%%%%%%%%%%%%%%%%%%%%%
% OLD
%First, we generate $T = 1000$ samples with inputs $x\in[-1,1]$ and outputs following a polynomial and sinusoidal relationship. 
%
% NEW
\rev{First, we generate a total of $T = 1000$ samples with input values $x\in[-2,2]$ and outputs following a polynomial and sinusoidal relationship. 
The dataset is split into three parts: (i) $500$ samples from the interval $x\in[-1,1]$, employed for training, (ii) $250$ samples from $x\in[1,2]$, which are used as a validation set for hyperparameter selection, and (iii) $250$ samples from $x\in[-2,-1]$, reserved as test set for performance evaluation.}
%%%%%%%%%%%%%%%%%%%%%%%%%%%%%%%%%%%%%%%%%%%%%%
Relying on \eqref{eqn:system.comp.lin}, we have
%$y = f_0(x) + f(x)^\top\bar\theta + \Delta(x) +e$, with 
\begin{equation}
    \begin{aligned}
    f_0(x) &= 0,\, f(x)^\top=[1, x, u, x^2\rev{,} u^2],\\
    \Delta(x) &= 0.7\sin(5x) + 0.5\cos(3x) + 0.4x^2 + 0.3x^3 \\&\quad\,\,- 0.2\sin(7x)\cos(2x),\\
\end{aligned}
\label{eqn:acad_sys}
\end{equation}
where $u = \sin(2\pi x) + 0.5 \cos(3\pi x)$ and $e$ follows a Gaussian distribution with standard deviation $0.1$. In the selected case study, we select $\bar\theta = [2,3,4,1.5,-0.8]$.
To improve the estimation accuracy in the presence of unknown nonlinearities $\Delta(x)$, we employ 
%a kernel-based model augmentation. Specifically, we use 
a Laplacian kernel function, where the kernel matrix $\mathbf{K}$ is computed as
$\rev{\mathbf{K}_{ij}} = \exp\left(-\frac{|x_i - x_j|}{\sigma}\right)$.%, with a bandwidth parameter $\sigma = 1$.

\begin{figure}
    \centering
    \includegraphics[trim={0cm 5.7cm 14.3cm 0cm}, clip=true, width=\linewidth]{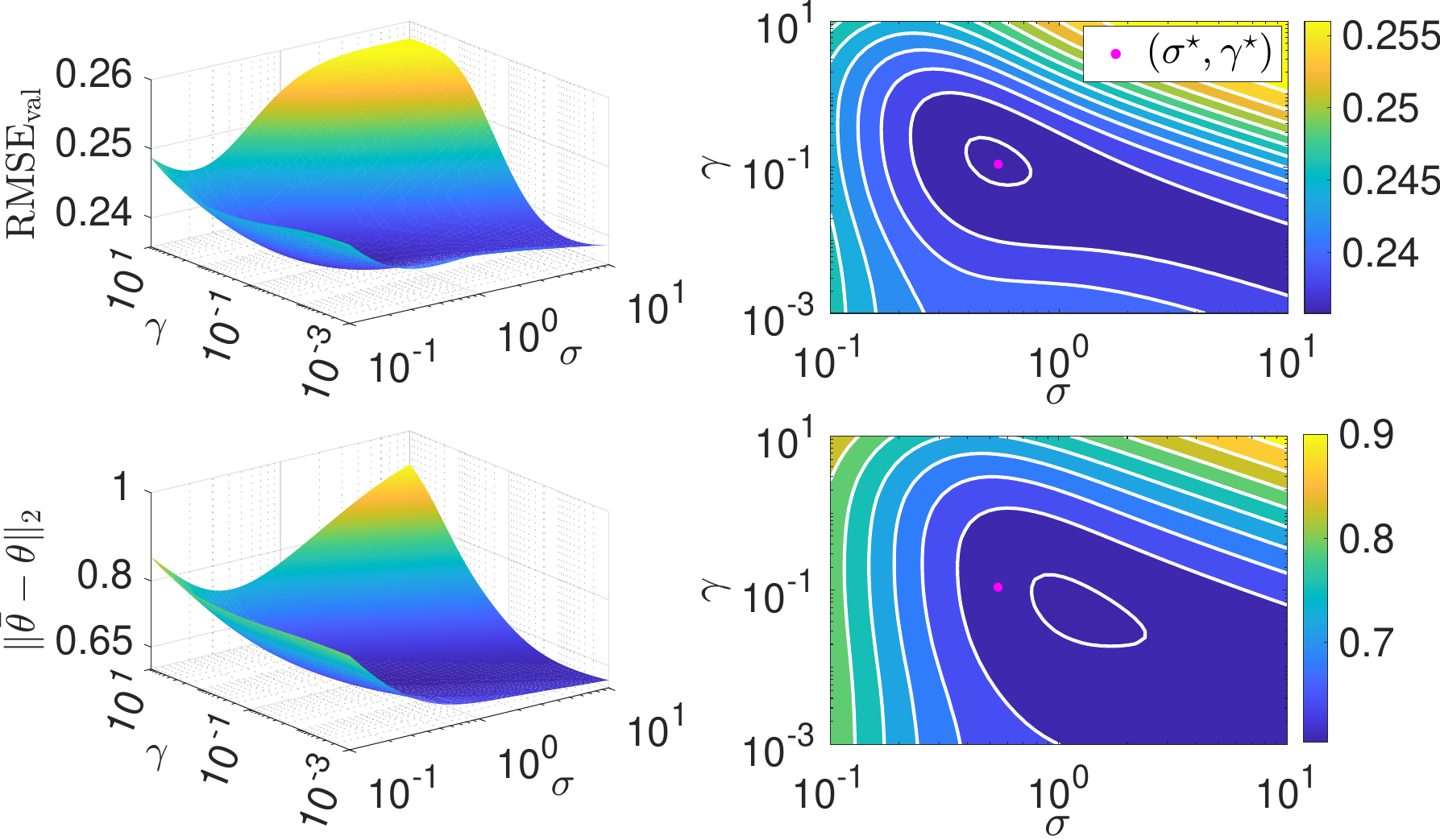}
    \caption{Validation RMSE as a function of the kernel bandwidth $\sigma$ and the regularization weight $\gamma$ (log scale), with the optimal hyperparameters $(\sigma^\star, \gamma^\star)$ (magenta dot) selected at the minimum RMSE region.}
    \label{fig:sigmagamma}
\end{figure}

\rev{To select the optimal hyperparameters $\sigma$ and $\gamma$, we rely on a validation-based procedure. Specifically, the training dataset is used to estimate the parameters for different combinations of $\sigma$ (kernel bandwidth) and $\gamma$ (regularization weight), while the validation dataset is employed to evaluate the root mean square error (RMSE) of the resulting models. The hyperparameters are tuned over a grid of $50\times50$ logarithmically spaced values for  $\sigma\in\left[10^{-1},10^{1}\right]$, and $\gamma\in\left[10^{-3},10^{1}\right]$, respectively, yielding a total of $2500$ candidate pairs. Among all tested combinations, the pair $(\sigma^\star, \gamma^\star) = (0.54, 0.11)$ is selected as it minimizes the RMSE on the validation dataset. The outcome of this procedure is illustrated in Fig.~1, which reports both the validation RMSE (RMSE$_\text{val}$) and the parametric error ($\|\bar\theta-\theta\|_2$) for each tested configuration of $\sigma$ and $\gamma$.
Notice that hyperparameters minimizing the RMSE do not necessarily coincide with those minimizing the parametric error. Indeed, being the true parameter vector $\bar{\theta}$ unknown in practice, it cannot be directly exploited. As shown in Fig.~\ref{fig:sigmagamma}, the RMSE-based selection nevertheless provides a reliable criterion, yielding parameter estimates that remain close to the minimum parametric error. }

\rev{To better illustrate the impact of $\gamma$ on the proposed identification process, we analyze its effect on the validation model performance for fixed $\sigma=\sigma^\star$. %In particular, we evaluate the RMSE on an independent validation set for different values of $\gamma$.
Figure~\ref{fig:gamma} depicts the RMSE on validation data as a function of $\gamma$, highlighting the trade-off between model flexibility and physical consistency. As expected, too small values of $\gamma$ lead to higher RMSE due to kernel overfitting to deviations and neglecting the physical priors. On the other hand, larger values result in biased parameter estimates as they enforce strict adherence to the physical model at the expense of not capturing relevant unmodeled terms. Clearly, achieving the correct trade-off between physical priors and the kernel-based representation of unmodeled terms is crucial for accurate identification.
\begin{figure}[!tb]
    \centering
    \includegraphics[width=0.9\linewidth]{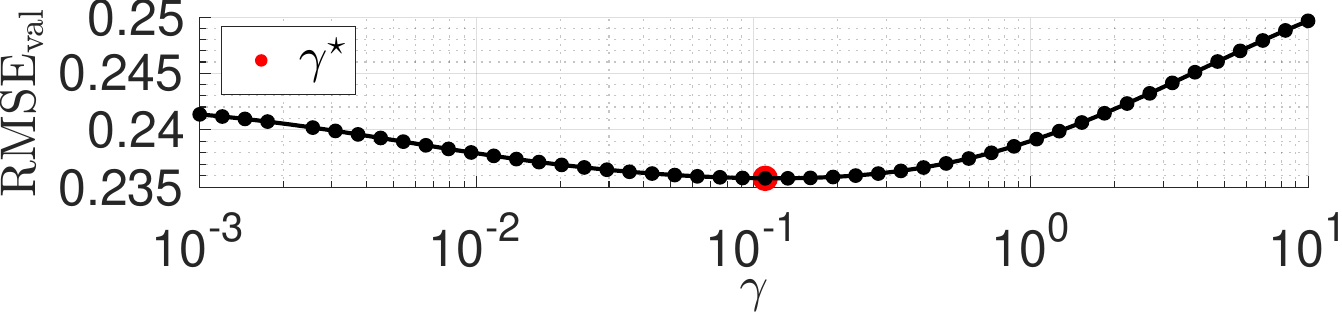}
    \caption{RMSE on validation data as a function of $\gamma$ (log scale).}
    \label{fig:gamma}
\end{figure}}

Once the hyperparameters are selected, the kernel-based estimator follows the formulation derived in Theorem \ref{thm:repthmII} and Theorem \ref{thm:retpthmII.lin}, where the correction is introduced through a kernel, and the operator $\Psi$ projects the observations into a feature space that captures unmodeled nonlinear effects.

{The obtained results are reported in Table \ref{tab:acex}, where the solution obtained with the proposed approach is compared with \rev{the true system (to provide easily comparable
information on the noise level),} the least-squares solution (LS), the discrepancy modeling approach (DM) described in the introduction (see, e.g.,~\cite{kaheman2019learning}), \rev{and a straightforward kernel ridge
regression (KRR) without the parametric part (i.e., prior knowledge on $f_0(x)$, $f(x)$ is not exploited)}\footnote{\rev{The same validation procedure was applied to tune the hyperparameters of both the KRR model and the DM approach, yielding $(\sigma^\star=0.1,\gamma^\star=10^{-2})$ for DM and $(\sigma^\star=10,\gamma^\star=10^{-2})$ for KRR, respectively.}}.} 
\rev{Results are compared in terms of obtained parameter estimates ($\theta_i$, $i = 1,\dots,5$), and RMSE on the test set (RMSE$_\text{tst}$), to assess the predictions accuracy.}

The estimated parameter vector $\theta^\star$ obtained using \eqref{eqn:thmlincf} with $(\sigma^\star,\gamma^\star)$ is significantly closer to the true $\bar\theta$ compared to the ordinary least-squares estimate $\theta^{LS}$, which does not account for unknown nonlinearities. The proposed kernel-based model achieves an RMSE of $0.343$, whereas the model obtained using $\theta^{LS}$, which neglects the unmodeled effects, results in a significantly higher RMSE of $0.961$.

{The estimate $\theta^{LS}$ also corresponds to the solution given by the discrepancy modeling approach. In this two-step procedure, the physical parameters are first estimated without accounting for $\Delta(x)$, and the resulting discrepancy is then modeled separately -- here, using a Laplacian kernel-based correction. While this approach reduces the RMSE to $0.929$, it falls short of the performance achieved by the proposed method, which simultaneously estimates both the physical parameters and the unmodeled dynamics. Moreover, it yields more biased parameter estimates. \rev{Notably, the performance of the standard KRR model confirms the importance of embedding prior physics. In fact, despite its flexibility, KRR does not estimate physical parameters and yields a substantially higher test RMSE, indicating overfitting to training data and limited generalization capability when used alone.}
\begin{table}[!tb]
    
    \centering
    \setlength{\tabcolsep}{3pt}
    \caption{Identification performance with different methods.}
    \begin{tabular}{|c|ccccc|c|}
    \hline
        & $\theta_1$ [-] & $\theta_2$ [-]& $\theta_3$ [-]& $\theta_4$ [-]& $\theta_5$ [-] & RMSE$_\text{tst}$ [-]\\
        \hline
        \textbf{True} & $2$ & $3$ & $4$ & $1.5$ & $-0.8$ & $0.097$\\
        \textbf{LS} &$2.55$ &$3.03$ &$4.32$ &$0.80$ &$-1.05$ & $0.961$\\
         \textbf{DM} &$2.52$ &$3.03$ &$4.32$ &$0.81$ &$-1.04$ & $0.929$\\
         {\textbf{KRR}} & -- & -- & -- & -- & -- & $1.206$\\
        \textbf{Proposed} & $\mathbf{2.18}$& $\mathbf{2.85}$&$\mathbf{4.16}$ &$\mathbf{1.20}$ &$\mathbf{-0.83}$ & $\mathbf{0.343}$\\
        \hline
    \end{tabular}
    \label{tab:acex}
\end{table}}

\begin{figure}
    \centering
    \includegraphics[width=\linewidth]{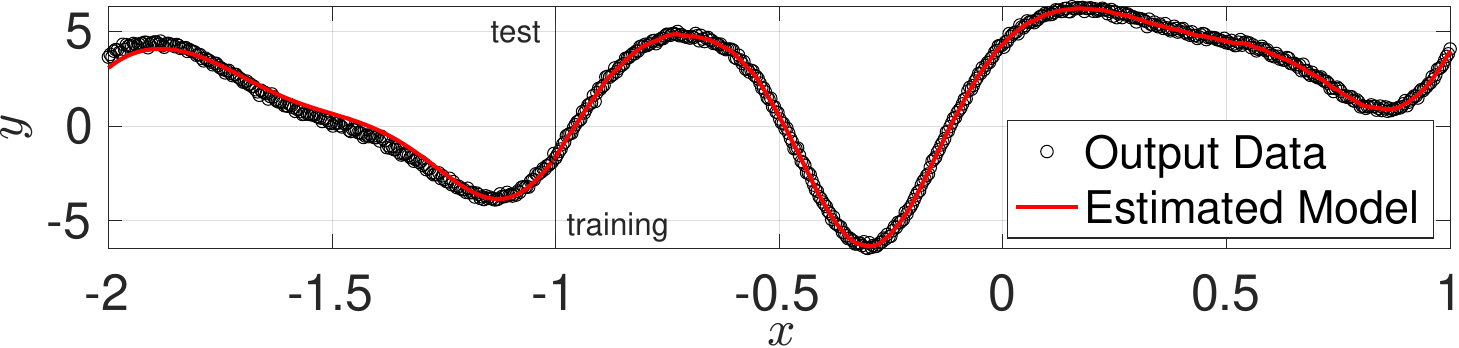}
    \caption{Estimated function and measured data for the test and training datasets.}
    \label{fig:output}
\end{figure}

\rev{Hence, figure~\ref{fig:output} illustrates a comparison between the estimated function and the measured data for the test and training datasets, confirming that the proposed kernel-based model accurately captures the nonlinear relationship underlying the dataset.}
{These results illustrate the benefits of using a kernel-based embedding approach. 
%The kernel correction allows the model to compensate for the unmodeled dynamics $\Delta$, effectively reducing the bias in the estimated parameters. In contrast, the least-squares solution relies solely on the known part of the model, leading to biased parameter estimates as it cannot account for the effects of~$\Delta$.}
%Moreover, the kernel-augmented model not only improves parameter estimation but also enhances prediction accuracy. 
By incorporating the unmodeled effects into the estimation process, the kernel-based approach better represents the system dynamics, leading to significantly lower RMSE values and improved parameter estimates.}
\begin{comment}
\textcolor{green}{
These findings highlight the effectiveness of our method in handling nonlinear discrepancies, demonstrating its potential for improving system identification in several applications.}
\GC{[Too much self-promotion in my opinion... besides, already stated at the beginning of the paragraph.]}
\end{comment}

{
To further assess the robustness and statistical significance of the obtained results, a Monte Carlo analysis is conducted over $1000$ independent identification experiments. 
In each run, a system following the structure in \eqref{eqn:acad_sys} is generated with randomly sampled noise realizations and true parameters generated around the nominal values 
$[2,3,4,1.5,-0.8]$ by adding uniform perturbations up to $\pm50\%$ of each component’s magnitude. 
The proposed kernel-based method (with Laplacian kernel) is again compared with a standard LS estimator relying only on the known structure, a DM approach obtained by first estimating the LS parameters and then identifying a Laplacian-kernel correction, and a pure kernel ridge regression KRR model 
without the physical component. For each trial, training, validation, and test sets were defined 
as before, and hyperparameters for the proposed, DM, and KRR approaches were selected through the same validation-based procedure. 

Table~\ref{tab:MC} reports the average parameter estimation error $\|\bar{\theta} - \hat{\theta}\|_2$, the fit percentage ($\mathrm{fit} \doteq 100\left( 1 - {\| y - \hat{y} \|_2}/{\| y - \bar{y} \|_2} \right) ,$ with $\bar{y} = \frac{1}{T}\sum_{t=1}^{T} y_t$) on the test 
data, and the test RMSE with their corresponding standard deviations over the $1000$ Monte Carlo runs. 
The results confirm the statistical consistency and robustness of the proposed approach, which 
achieves the smallest mean parameter error and the lowest test RMSE, with limited variability 
across trials.}
\begin{table}[t]
\centering

\caption{Statistical evaluation over $1000$ Monte Carlo experiments. Mean $\pm$ standard deviation are reported.}
\label{tab:MC}
\begin{tabular}{|c|ccc|}
\hline
& $\|\bar{\theta} - \hat{\theta}\|_2$ [-]& {fit$_\text{tst}$ [\%]} & {RMSE}$_{\text{tst}}$ [-] \\
\hline
\textbf{True} & $0\pm0$ & $96.07 \pm 1.45$ & $0.099 \pm 0.0045$ \\
\textbf{LS} & $0.966 \pm 0.014$ & $62.44 \pm 13.83$ & $0.953 \pm 0.032$ \\
\textbf{DM} & $0.966 \pm 0.014$ & $64.91 \pm 13.34$ & $0.891 \pm 0.104$ \\
\textbf{KRR} & -- & $38.80 \pm 12.16$ & $1.281 \pm 0.241$ \\
\textbf{Proposed} & $\mathbf{0.517 \pm 0.056}$ & $\mathbf{78.74 \pm 7.67}$ & $\mathbf{0.541 \pm 0.138}$ \\
\hline
\end{tabular}
\end{table}

\subsection{Cascade tank system benchmark}
In the second example, we test the proposed approach on a CTS benchmark \cite{schoukens2016cascaded} and compare it with other state-of-the-art
estimation methods. 
The CTS regulates fluid levels among two connected tanks and a pump. First, water is pumped into the upper tank, then it flows to the lower tank and back to the reservoir. Overflow occurs when a tank is full: The excess from the upper tank partially drains into the lower one, with the rest exiting the system. The water level is measured using a capacitive water level sensor.
In \cite{schoukens2016cascaded} an approximate nonlinear, continuous state-space model of the CTS is derived using Bernoulli's and mass conservation principles. {Here, we rely on its discretized version, i.e.,}
\begin{equation}
\begin{aligned}
    x_{1,t+1} &= x_{1,t} + T_s\left(-k_1 \sqrt{x_{1,t}} + k_4 u_t + e_{1,t}\right),\\
    x_{2,t+1} &= x_{2,t} + T_s\left( k_2 \sqrt{x_{1,t}} - k_3\sqrt{x_{2,t}} + e_{2,t}\right),\\
    y_t &= x_{2,t} + w_t,
\end{aligned}
\label{eqn:tctmodel}
\end{equation}
where $u_t \in \mathbb R$ is the input signal, $x_{1,k}\in \mathbb R$ and $x_{2,k}\in \mathbb R$ are the states of the system, $y_t \in \mathbb R$ is the output, and $e_{t}\in \mathbb R^2$, $w_t\in \mathbb R$ are the additive noise sources. The sampling time is set to $T_s = 4 s$. Moreover, the system is characterized by four unknown physical constants, $k_1$, $k_2$, $k_3$, and $k_4$, which depend on the properties of the system and need to be estimated. Since this model ignores the water overflow effect, and the water level sensors introduce an extra source of nonlinear behavior, unmodeled dynamics are included in the physical dynamics model \eqref{eqn:tctmodel}. 
The training and validation datasets consist of $T = 1024$ input-output samples. \rev{A portion of the $10\%$ of the training data was reserved for hyperparameter tuning, carried out as in the previous example, while the original benchmark validation dataset was kept unchanged to ensure a fair comparison with existing results.}

The goal is to estimate the dynamics of the system using only the available training data. For the identification,  we employ the nonlinear state smoothing, described by Algorithm~\ref{alg:kernel_state_space},  assuming that $f(x_t,u_t,\theta)$ and $g(x_t,u_t,\theta)$ are defined according to the discretized model \eqref{eqn:tctmodel} and selecting $P_e=10^{-3}I_2$, $P_w=10^{-2}$, $P_0=0.5 I_2$, $\theta_0= [0.05, 0.05, 0.05, 0.05]^\top$, and $\hat x_0=[y_0,y_0]^\top$. Moreover, we set the UKF weights 
%$a$, $w^m$, $w^c$ 
according to the formulation in \cite{wan2000unscented}, that is $a=2.74$, $w^m_0 = 0.33$, $w^c_0 = 2.33$, and $w^m_i = w^c_i = 0.67$, for $i=1,\dots,2n$.
Then, we solve an optimization problem of the form \eqref{eqn:optprob.theta} for $\gamma=0.1$. Specifically, once the smoothed trajectory of the unmeasured state $\hat{x}_{1,0:T-1}^s$ is computed, we define a predictive model of the form \eqref{eqn:predmodel.new} and we solve Problem~\eqref{eqn:optprob.theta} applying Theorem~\ref{thm:repthmII}. Specifically, we employ the \textit{fmincon} solver using a \textit{sqp} method to solve Problem~\eqref{eqn:reprthm.ext1}. Hence, considering \eqref{eqn:tctmodel} and selecting $z_t \doteq [\hat{x}_{1,t-1}^s, y_t, u_{t-1}]$, we define $\xi(z_t) \doteq y_{t} + T_sk_2 \sqrt{\hat x^s_{1,t-1} + T_s\left(-k_1 \sqrt{\hat x^s_{1,t-1}} + k_4 u_{t-1}\right)} - T_sk_3\sqrt{y_{t}}$. Thus, once we obtain $(\theta^\star,\delta^\star)$ as the solution of \eqref{eqn:optprob.theta} according to Theorem \ref{thm:repthmII}, the optimal estimation model becomes $\hat y_{t+1} = \xi(z_t,\theta^\star) + \delta^\star(z_t)$, selecting the Gaussian kernel $\kappa(x, x^{\prime}) = \exp(-\|x - x^{\prime}\|^2 / 2\sigma^2)$ with $\sigma=11$.

To evaluate the effectiveness of the estimation algorithm, as suggested in \cite{schoukens2016cascaded}, the RMSE is employed as the performance metric. Table \ref{tab:comparison} presents the performance of the proposed kernel-based identification method compared with the solution obtained by solving \eqref{eqn:optprob.theta} when no kernel is used to compensate for unmodeled dynamics in \eqref{eqn:tctmodel}. The results are reported for the estimation and validation datasets, considering: (i) the \textit{prediction} task, i.e., given $z_t$, we estimate $y_{t+1}$, and (ii) the \textit{simulation} task, i.e., we recursively estimate $y_{t+1}$ by defining $z_t$ with $\hat y_t$ (the previous estimate of $y_t$). Moreover, for the simulation results, we also report the fit values in Table~\ref{tab:comparison}. These first results highlight the significant reduction in the RMSE achieved through kernel embedding, particularly in the simulation setting. Notably, despite the optimization being performed minimizing the prediction error, the joint use of the kernel-based approach and the smoother substantially improves the multi-step simulation accuracy. 
\begin{comment}
\begin{table}[!tb]
    \centering
    \caption{RMSE performance on estimation and validation data.}
    \begin{tabular}{|l|cc|cc|}
        \hline
                      & \multicolumn{2}{c|}{\textbf{Prediction}} & \multicolumn{2}{c|}{\textbf{Simulation}} \\
                      & Train. & Val. & Train. & Val. \\
        \hline
        \textbf{Without Kernel}   & 0.069 & 0.075 & 0.557 & 0.617\\
        \textbf{With Kernel} & 0.030 & 0.049 & 0.074 & 0.306\\
        \hline
    \end{tabular}
    \label{tab:comparison}
\end{table}
\end{comment}
\begin{table}[!tb]
    \centering
    \caption{Performance analysis on estimation and validation  data.}
    \begin{tabular}{|l|cc||cc|}
        \hline
                      & \multicolumn{2}{c||}{\textbf{Pred.} (RMSE [V])} & \multicolumn{2}{c|}{\textbf{Sim.} (RMSE [V], fit [\%])} \\
                      & Train. & Val. & Train. & Val. \\
        \hline
        \textbf{\eqref{eqn:tctmodel} only}   & $0.06$ & $0.06$ & $0.37$, $83.14$ & $0.37$, $82.46$\\
        \textbf{\eqref{eqn:tctmodel} + kernel} & $0.04$ & $0.05$ & $0.17$, $92.11$  & $0.18$, $91.55$\\
        \hline
    \end{tabular}
    \label{tab:comparison}
\end{table}
This improvement is also reflected in the results reported in Table \ref{tab:comp_tanks}, where we compare the simulation performance for the validation data with other state-of-the-art approaches from the literature. Among these methods, we observe that the proposed approach also outperforms the solution of our multi-step identification approach presented in \cite{AutomaticaExtendedVersion}. {This improvement is likely due to the ability of the kernel-based method to automatically select an optimal representation for the approximating term $\delta(z_t)$, whereas in \cite{AutomaticaExtendedVersion} this term was chosen from a limited, predefined dictionary of basis functions.}

Finally, we report the identified parameters for completeness. The smoothed initial condition is $\hat{x}^s_0 = [4.78, 5.20]$ whereas, for $\hat{k}$, we have: (i) with no kernel, $\hat{k} = [-0.01, 0.05, 0.06, 0.01]$, and (ii) with kernel, $\hat{k} = [0.08, 0.05, 0.05, 0.06]$. 
\begin{table}[!tb]
    \centering
    \caption{Methods comparison (simulation RMSE on validation data).}
    \renewcommand{\arraystretch}{0.98} % Reduce row height
    \begin{tabular}{|lc||lc|}
    \hline
        \textbf{Method} & \textbf{RMSE} [V] & \textbf{Method} & \textbf{RMSE} [V] \\
    \hline
        Svensson et al. \cite{svensson2017flexible} & $0.45$ & PWARX \cite{mattsson2018identification} & $0.35$ \\
        Volt.FB \cite{schoukens2016modeling} & $0.39$ & SED-MPK \cite{dalla2021kernel} & $0.48$ \\
        INN \cite{mavkov2020integrated} & $0.41$ & PNLSS-I \cite{relan2017unstructured} & $0.45$ \\
        NLSS2 \cite{relan2017unstructured} & $0.34$ & NOMAD \cite{brunot2017continuous} & $0.37$ \\
    \hline
        \textbf{Donati et al.} \cite{AutomaticaExtendedVersion} & $0.26$ & \textbf{Proposed} & $\mathbf{0.178}$ \\
    \hline
    \end{tabular}
    \label{tab:comp_tanks}
\end{table}

%\vspace{-0.5cm}
\section{Conclusions}\label{sec:concl}
This work introduced a kernel-based framework for physics-informed nonlinear system identification, effectively embedding
%data-driven augmentation. \CD{New:
kernel methods
%instead of data-driven augmentation}
%\GC{[Again, I do not understand what we mean by data-driven augmentation. All identification methods are somehow data-driven, since we use input/output data to identify the model...]}
%
while preserving physical model interpretability. 
%\textcolor{green}{By incorporating nonlinear state smoothing, we addressed cases with unobservable states, achieving superior prediction and simulation accuracy. The numerical results confirm the effectiveness of kernel augmentation in compensating for unmodeled dynamics.}
%
{Then, to tackle the case of unmeasured states, we incorporate a
nonlinear state smoothing. The numerical results confirm the effectiveness of the proposed approach in finding meaningful parametric models with compensating unstructured components.}

\rev{
Future investigations will explore advanced strategies for exploiting state smoothing with identification, analyzing the role of model uncertainty. Clearly, the use of state estimates rather than the true (unknown) states may introduce small bias and increased variance. Using URTSS-smoothed estimates significantly reduces such effects, but further improvements can be obtained.
Potential directions include weighted fitting based on the smoother covariance $P^s_t$, iterative refinement of the filtering model, and the tuning of noise covariances to account for modeling inaccuracies, as well as the derivation of error bounds in the smoothing step.}
\rev{Additional future work may focus on improving the hyperparameter selection procedure, for instance, by incorporating external physical information or constraints that the parameters must satisfy, which could guide the choice of $\gamma$ and ultimately reduce the parametric error.}

\section{Acknowledgments}
The authors are grateful to T.\ Alamo for his valuable suggestion to incorporate kernel-models in our system identification framework.

\bibliographystyle{IEEEtran}
\bibliography{root.bib}

\end{document}